\documentclass[aps,prl,superscriptaddress,epsfigm,twocolumn,showpacs]{revtex4}
\usepackage{graphicx}
\usepackage{amsmath}

\begin{document}

\title{Triplet-Singlet Spin Relaxation in Quantum Dots with Spin-Orbit Coupling}

\author{Juan I. Climente}
\email{climente@unimore.it} \homepage{www.nanoscience.unimore.it}
\affiliation{CNR-INFM S3, Via Campi 213/A, 41100 Modena, Italy}
\author{Andrea Bertoni}
\affiliation{CNR-INFM S3, Via Campi 213/A, 41100 Modena, Italy}
\author{Guido Goldoni}
\affiliation{CNR-INFM S3, Via Campi 213/A, 41100 Modena, Italy}
\affiliation{Dipartimento di Fisica, Universit\`a degli Studi di Modena e Reggio Emilia,
Via Campi 213/A, 41100 Modena, Italy}
\author{Massimo Rontani}
\affiliation{CNR-INFM S3, Via Campi 213/A, 41100 Modena, Italy}
\author{Elisa Molinari}
\affiliation{CNR-INFM S3, Via Campi 213/A, 41100 Modena, Italy}
\affiliation{Dipartimento di Fisica, Universit\`a degli Studi di Modena e Reggio Emilia,
Via Campi 213/A, 41100 Modena, Italy}
\date{\today}

\begin{abstract}

We estimate the triplet-singlet relaxation rate due to spin-orbit
coupling assisted by phonon emission in weakly-confined quantum
dots. Our results for two and four electrons show that the different
triplet-singlet relaxation trends observed in recent experiments
under magnetic fields can be understood within a unified theoretical
description, as the result of the competition between spin-orbit
coupling and phonon emission efficiency. Moreover, we show that both
effects are greatly affected by the strength of the confinement and
the external magnetic field, which may give access to very
long-lived triplet states as well as to selective population of
the triplet Zeeman sublevels.


\end{abstract}

\pacs{73.21.La,71.70.Ej,72.10.Di,73.22.Lp}

\maketitle



Semiconductor quantum dots (QDs) are called to play a central role in the
emerging field of spintronics, because their zero-dimensional confinement
constitutes an optimal environment to manipulate the spin of bound
electrons.\cite{ZuticRMPAtatureSCI}
This has stimulated their use as spin-filters\cite{CiorgaPRB}, as well as
several attempts to realize solid state implementations of
spin-based qubits.\cite{CoishPSS} Understanding the spin relaxation
in these structures is of utmost interest for their eventual use in
practical devices. This has triggered a large number of experimental
works in the last few years, where two main classes of spin
transitions have been investigated, namely the spin-flip between
single-electron Zeeman sublevels\cite{CoishPSS} and the
triplet-singlet (TS) transition in QDs with an even number of
electrons.\cite{FujisawaNATJPCM,SasakiPRL,HansonPRL, PettaSCI}
Remarkably, while the former systems have received much theoretical
attention\cite{CoishPSS,StanoPRB}, the understanding of the latter
is still rather limited. In particular, the TS relaxation due to
spin-orbit (SO) coupling -- which is often the dominant
spin
relaxation mechanism in semiconductor
QDs -- has only briefly been addressed in two-electron
QDs\cite{FlorescuPEPRB,ChaneyPRB}, and many relevant features
observed in experiments remain uncomprehended.
This is the case of the role of an external magnetic field:
experimental measuraments away from the TS anticrossings suggest that
the influence of axial fields on the spin relaxation is fairly weak\cite{FujisawaNATJPCM,SasakiPRL}.
This is in strong contrast with the single-electron case, where a power-dependence
of the relaxation rate on the field has been demonstrated\cite{CoishPSS},
and it might seem surprising because the field reduces the energy splitting between
the singlet and the triplet, what should enhance SO coupling. 
Besides, in the vicinity of the TS anticrossing, both
decreased\cite{SasakiPRL} and increased\cite{HansonPRL} relaxation
rates have been reported. In this context, a unified picture
describing the effect of an axial magnetic field on the TS spin
relaxation rate is on demand.

In this Letter, we study the TS spin relaxation due to SO coupling
in circular QDs with weak lateral confinement. Acoustic phonon
emission, assisted by SO interaction, has been shown to be the
dominant relaxation mechanism in this kind of QDs when cotunneling
and nuclei-mediated relaxation are reduced. In fact, cotunneling is
an extrinsic scattering process and can be controlled by means of
reduced tunneling rates\cite{SasakiPRL}, while nuclei-mediated
relaxation dramatically decreases with external magnetic
fields.\cite{CoishPSS}
We show that the current experimental evidence
\cite{FujisawaNATJPCM,SasakiPRL,HansonPRL} can be reconciled within
a unified picture, where the field dependence of the relaxation rate
is determined by the interplay between spin-orbit coupling
and phonon emission efficiency. Furthermore, we show that such
interplay can be tailored in order to obtain improved spin
lifetimes.

In weakly confined QDs, correlation effects may strongly influence
charge and spin excitations.\cite{GarciaPRL} In order to calculate
the relaxation time of excited few-electron correlated states, we
need to know both ground and excited states with comparable accuracy:
our method of choice is the full configuration interaction (FCI).\cite{RontaniJPC}
The single-electron states are calculated within the effective mass
approximation for a typical ``vertical'' GaAs/AlGaAs QD,
with confinement potential
$V(\mathbf{r})=V_z(z) + 1/2\,m^*\,\omega_0^2\,(x^2+y^2)$,
$V_z(z)$ representing the (finite) vertical confinement of a quantum
well of thickness $W$, $\hbar \omega_0$ being the single-electron energy
spacing of a lateral two-dimensional harmonic trap, and $m^*$ the
effective mass. The lateral confinement is much weaker than the vertical one,
and a magnetic field, $B$, is applied along $z$.
Under these conditions, the low-lying single-electron states are well described
by the Fock-Darwin spectrum and the lowest eigenstate of the quantum
well.\cite{ReimannRMP}
The single-electron levels can be classified by their radial quantum number
$n=0,1\ldots$, azimuthal angular momentum $m=0,\pm 1\ldots$, and spin
$s_z=\uparrow, \downarrow$. In turn, the few-electron states can be labelled
by the total azimuthal angular momentum $M$,
total spin $S$, its $z$-projection $S_z$, and by the number $N=0,1,\ldots$ 
indexing the energy order.

We introduce the SO coupling via the linear Rashba and Dresselhaus
terms, 
${\cal H}^R$ and ${\cal H}^D$ respectively. For a quantum well grown
along the $[0 0 1]$ direction, these terms can be written as:\cite{RashbaDressCartoixaPRB}
\begin{equation}
\label{eq1}
{\cal H}^R = \frac{\alpha}{i \hbar} (\pi^+ s_z^- - \pi^- s_z^+); \;
{\cal H}^D = \frac{\beta}{\hbar} (\pi^+ s_z^+ + \pi^- s_z^-),
\end{equation}

\noindent where $\alpha$ and $\beta$ are the Rashba and Dresselhaus coefficients
for the sample under study. $\pi^{\pm}$ and $s_z^{\pm}$ are ladder operators,
which change $m$ and $s_z$ by one unit,
respectively.
%
Since the few-electron $M$ and $S_z$ quantum numbers are given by the algebraic
sum of their single-electron counterparts, Rashba interaction mixes
$(M,\,S_z)$ states with $(M\pm 1,\,S_z\mp 1)$ ones,
and Dresselhaus interaction mixes $(M,\,S_z)$ states with
$(M\pm 1,\,S_z\pm 1)$ ones.


In our calculation, the
SO terms of Eq.~(\ref{eq1}) are diagonalized in a basis of few-electron states,
which are
computed as linear combinations of Slater determinants,
according to the FCI method.\cite{RontaniJPC,details_calc}
In the general case the Rashba and Dresselhaus terms
break $S$ and $M$ symmetries.
However, for GaAs QDs, SO coupling is but a small perturbation and
the quantum numbers $M$, $S$ and $S_z$ are approximately valid
except in the vicinity of the anticrossing
regions.\cite{FlorescuPEPRB,PietilainenPRB} Thus, we will still use
them for clarity of the discussion. We estimate the relaxation rate
at zero temperature due to acoustic phonon emission. The
electron-phonon interaction is taken into account as in
Ref.~\onlinecite{ClimentePRB}. Hence, we consider not only
deformation potential as in previous works
(Ref.~\onlinecite{FlorescuPEPRB}), but also piezoelectric field
scattering. The piezoelectric field interaction is dominant when the
phonon energy is small,\cite{ClimentePRB,StanoPRB} so that it
provides the main contribution to the relaxation in the interesting
regions of TS anticrossings. GaAs material parameters
are taken in the calculations,\cite{ClimentePRB}
along with a Land\'e factor $g\!=\!-0.44$.

\begin{figure}[h]
\includegraphics[width=8cm]{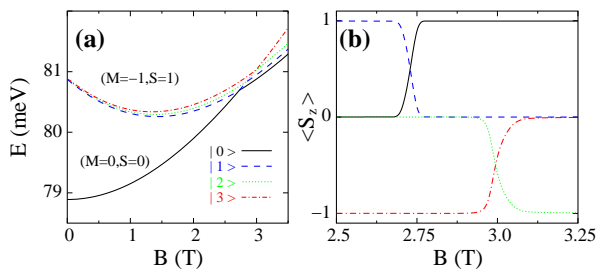}
\caption{(Color online). (a) Energy of the four lowest-lying states
$|N\rangle$ of a two-electron QD with $W=10$ nm and $\hbar
\omega_0=4$ meV as a function of the magnetic field. The approximate
quantum numbers $(M,S)$ are shown.
(b) $S_z$ expectation value of the four lowest
energy levels in the region of the singlet-triplet
anticrossing.}\label{Fig1}
\end{figure}

We start our discussion with the two-electron case (see
Fig.~\ref{Fig1}). We use a typical value of the Dresselhaus
coefficient for a GaAs QD, $\beta\!=\!25$ meV$\cdot$\AA, and a
Rashba coefficient $\alpha\!=\!5$ meV$\cdot$\AA, which could be
ascribed e.g. to a small accidental asymmetry of the quantum well.
The low-lying singlet state and the excited triplet state with three
Zeeman sublevels are shown in Fig.~\ref{Fig1}(a). With increasing
magnetic field, the singlet anticrosses with all triplet sublevels.
The anticrossing energy gap is very small (of the order of $\mu$eV),
as expected for GaAs QDs\cite{FujisawaNATJPCM}, and it is
particularly small for the $S_z\!=\!0$ triplet sublevel. This is
because the Dresselhaus (Rashba) interaction mixes the singlet with
the triplet $S_z\!=\!-1\,(+1)$ sublevel, but does not mix states
with $\Delta S_z=0$, which, therefore, takes place only indirectly
through higher-lying states. Figure \ref{Fig1}(b) illustrates the
expectation value of $\langle S_z \rangle$ of the four lowest-lying
levels around the TS anticrossing. One can see that SO interaction
barely affects the spin quantum numbers except in a narrow magnetic
field range around the anticrossings.\cite{FlorescuPEPRB}\\

\begin{figure}[h]
\includegraphics[width=7.5cm]{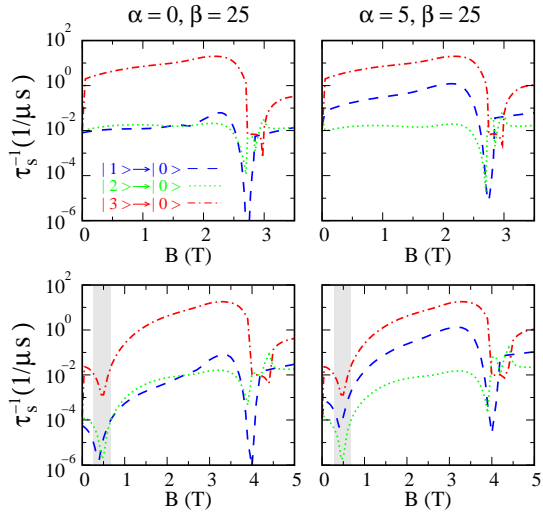}
\caption{(Color online). Spin relaxation rate of the three first
excited states in two-electron QDs vs.~magnetic field. Top row:
$W=10$ nm, $\hbar \omega_0=4$ meV; bottom row: $W=14$ nm, $\hbar
\omega_0=5$ meV. The SO interaction coefficients are shown on top of
each column (units of meV$\cdot$\AA). Note that before the
anticrossing dip, $|N\rangle \!=\! 1,2,3$ correspond to the triplet
sublevels $S_z\!=\!+1,0,-1$, respectively. In the bottom panels, the
shaded areas highlight the magnetic field window of
geometrically-suppressed phonon emission (see text).}\label{Fig2}
\end{figure}

In Fig.~\ref{Fig2} we analyze the relaxation rate from the three
first excited to the ground state of two-electron QDs with
different dimensions.
Left (right) panels correspond to structures without (with) Rashba
interaction.
For the QD studied in the upper panels, one can see that the relaxation
rate increases slowly with the magnetic field, and then it suddenly
drops in the anticrossing region ($B \sim 2.25 - 3.25$ T).\cite{a0T}
This behavior,
whose physical mechanism will be explained in the following,
is in qualitative agreement with various recent
experiments in weakly-confined QDs.
In particular
the rather weak dependence with the field before the anticrossing
agrees well with TS relaxation measuraments\cite{FujisawaNATJPCM,SasakiPRL},
the increased relaxation rate before the anticrossing has been
reported in Ref.~\onlinecite{HansonPRL}, and the reduced rate
in the anticrossing region may be inferred from the long triplet
lifetimes for eight-electron QDs with small
ST energy splittings\cite{SasakiPRL}.

The general trends described above can be explained by the opposite
effect of the magnetic field on the SO mixing and the phonon
emission efficiency. On the one hand, as the singlet-triplet energy
splitting decreases, SO interaction couples the states more
efficiently, favouring spin relaxation. On the other hand, the
phonon energy decreases, reducing the efficiency of the
electron-phonon interaction. The latter effect, which follows from
the different orbital quantum numbers of the initial and final
electron states, occurs at a rate that is determined by the
interplay between the acoustic phonon wavelength and the dimensions
of the QD.\cite{BockelmannPRB,ClimentePRB,StanoPRB} For the QD of
the upper panel in Fig.~\ref{Fig2}, the effect of the magnetic field
on the SO interaction and phonon emission is mostly of similar
magnitude, which explains the weak changes of the relaxation rate.
At the anticrossing point, in spite of the fact that the SO mixing
is maximum [see Fig.~\ref{Fig1}(b)], the phonon energy is so small
(few $\mu$eV) that the spin relaxation is strongly supressed.
It is worth mentioning that this result is opposed to that predicted
for the TS anticrossing in a lateral QD, where maximum relaxation
rate is predicted at the anticrossing point.\cite{FlorescuPEPRB} The
origin of this difference might lie on the fact that the
electron-phonon interaction matrix elements of lateral QDs with
strongly asymmetric (non-parabolic) confinement potential may be
significant even for very small phonon energies.\cite{Marian} As a
result, SO interaction alone would dominate spin relaxation in such
structures.

As stated before, the dimensions of the QD are known to play a critical
role to determine the phonon emission
efficiency.\cite{BockelmannPRB,ClimentePRB,StanoPRB}
To illustrate this effect
on the spin relaxation, in the bottom panels of
Fig.~\ref{Fig2} we report the rate of a QD with larger height and stronger
lateral confinement than the previous one. As compared to the upper panels,
one can see a visibly stronger dependence of the relaxation rate on the field.
This is because for the present QD dimensions, the phonon emission
efficiency turns out not to be balanced with the SO mixing effect.
As a result, when the effect of SO coupling prevails the relaxation rate
strongly increases with $B$. Still, in the anticrossing region,
that is now shifted towards higher magnetic fields,
the suppression
of phonon emission again yields a relaxation minimum.
Furthermore, a new feature emerges in this case,
namely a dip at about $B\sim 0.5$ T (shaded area in the panels)
This dip, which we had previously predicted for charge relaxation in
QDs,\cite{ClimentePRB}
comes from the geometrically-induced
suppression of the phonon emission,
occuring when the quantum well width
is a multiple of the phonon wavelength $z-$projection.
This feature may give access to very long-lived triplet states at
finite values of the magnetic field\cite{preChaneyPRB}, where
phonon-induced relaxation is usually the dominant scattering mechanism.
Moreover, this dip takes place at a $B$ value where the initial and final
electron states are well-resolved energetically, which renders this minimum
more useful than the one coming from the TS anticrossing.
In the illustrated case, the average relaxation rate of the three Zeeman
sublevels in this dip corresponds to a triplet relaxation time of tenths
of seconds, two orders of magnitude over the longest triplet
lifetime reported to date.\cite{HansonPRL}
The position and depth of this kind of relaxation minima depend on the QD height
and the emitted phonon energy. Therefore, they are almost independent of
the SO interaction in the structure, which in GaAs has a neglegible
influence on the phonon energy.
%
%
%

Next, we focus on the effect of the separate Rashba and Dresselhaus
contributions over the spin relaxation by comparing the left and
right panels of Fig.~\ref{Fig2} in the magnetic field region before
the TS anticrossing. When only Dresselhaus terms are present (left
panels), the singlet mixes directly only with the higher-lying
($S_z\!=\!-1$) Zeeman sublevel of the triplet. As a result,
relaxation from such Zeeman sublevel (dot-dashed line) is about two
orders of magnitude faster than from the $S_z\!=\!0,+1$ sublevels,
 and it exhibits a stronger dependence on the field.
When a small Rashba interaction is switched on (right panels),
 direct mixing of the singlet with the triplet $S_z\!=\!+1$ sublevel
is enabled. This accelerates the relaxation rate from this
sublevel (dashed line) in one order of magnitude 
 and introduces a stronger dependence on $B$.
It is worth noting that the order-of-magnitude enhancement
of the relaxation rate due to the Rashba interaction is present away
from the anticrossing region, where the effect of the SO interaction
on $\langle S_z \rangle$ is barely visible [see Fig.~\ref{Fig1}(b)].
From the above discussion it follows that
in a magnetic field both Rashba and Dresselhaus
interactions play an important role in determining the
TS spin relaxation rate,
as opposed to the well-known single-electron
case, where the relaxation is mostly due to Rashba coupling only.\cite{CoishPSS}
Moreover, we see that the lifetimes of the triplet Zeeman sublevels
may strongly differ depending on the relative Rashba and Dresselhaus
contributions. This may be useful to selectively populate the triplet
sublevels.\\

\begin{figure}[h]
\includegraphics[width=6.5cm]{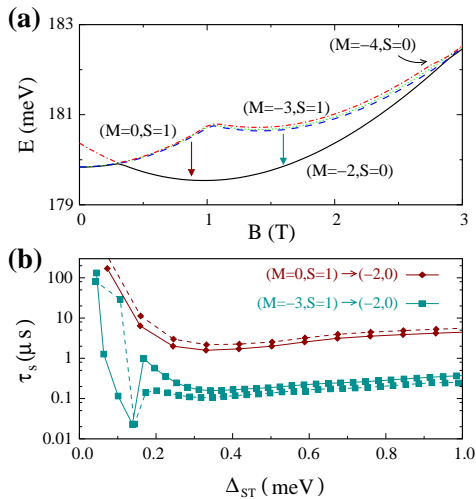}
\caption{(Color online). (a) Energy of the four lowest-lying levels
of a four-electron QD as a function of the magnetic field.
The QD has $W\!=\!10$ nm and $\hbar \omega_0\!=\!4$ meV,
$\alpha\!=\!15$ meV$\cdot$ \AA~ and $\beta\!=\!25$ meV$\cdot$ \AA.
The approximate quantum numbers $(M,S)$ are shown.
Arrows are used to indicate the two different spin
transitions we compare.
(b) Average triplet lifetime for the two spin transitions
we compare, as a function of the singlet-triplet energy
splitting. $\beta\!=\!25$ meV$\cdot$\AA, and solid (dashed) lines
are used for $\alpha\!=\!0$ ($\alpha\!=\!15$) meV$\cdot$\AA.}\label{Fig3}
\end{figure}

We now investigate the TS spin relaxation in a
four-electron QD. The energy spectrum of the lowest-lying triplet
and singlet states in a magnetic field, plotted in Fig.~\ref{Fig3}(a),
 is very different from that of the two-electron case, but it
closely resembles the one found experimentally for eight-electron QDs in
Ref.\onlinecite{SasakiPRL} (except for the absence
of eccentricity features in the zero field limit\cite{TokuraPB}).

Here, we investigate the spin relaxation rate in the region
$B \sim 0.3 - 3$ T, where the ground state is a singlet ($M\!=\!-2$) and
the first excited state is a triplet with two possible values of
the angular momentum, depending on the magnetic field:
for $B \!<\! 1$ T the angular momentum is $M\!=\!0$, and for $B \!>\! 1$ T
it is $M\!=\!-3$.
These states are well separated from higher-lying states, so they
might be used as a two-level system for quantum computation
purposes.
We compare the lifetimes of both triplet states in
Fig.~\ref{Fig3}(b), where the averaged lifetimes of
the three Zeeman sublevels are plotted as a function
of the singlet-triplet energy splitting, $\Delta_{ST}$,
with (dashed lines) and without (solid lines) Rashba interaction
(Dresselhaus interaction is present in both cases).
 The qualitative behavior is similar for both states: the lifetime
is roughly constant for large energy splittings
($\Delta_{ST}>0.25$ meV), and it increases when the energy splitting
is small ($\Delta_{ST}<0.25$ meV). This behavior, which is in agreement with
the experimental findings of Ref.~\onlinecite{SasakiPRL}, can be understood
in the same terms of compensation between SO coupling and phonon emission
efficiency as in the two-electron cases studied above.
The strong dip of the $M\!=\!-3$ triplet at $\Delta_{ST} \sim 0.15$ meV
is due to the anticrossing of the upper Zeeman sublevel with the $M\!=\!-4$ singlet
at strong magnetic fields [at $B \sim 2.8$ T in Fig.~\ref{Fig3}(a)], which strongly
enhances spin relaxation. For smaller $\Delta_{ST}$, though, the small
phonon energy again leads to increased lifetimes.

An important result shown in Fig.~\ref{Fig3}(b) is that the average
lifetime of the triplets differs by over one order of magnitude
dependening on their angular momentum, regardless of the (here
fairly strong) Rashba interaction. This is because the $M\!=\!0$
triplet differs from the $M\!=\!-2$ singlet ground state in two
quanta of angular momentum, and therefore direct SO mixing is not
possible. In contrast, direct mixing is possible for the $M\!=\!-3$
triplet, and this makes the relaxation much faster. It then follows
that, by using four-electron QDs instead of two-electron ones, one
can use an external magnetic field to select excited states whose
spin transition is ``forbidden'' even in the presence of linear SO
interaction. This result is consistent with recent measuraments,
where different lifetimes were observed for triplet states with
different orbital quantum numbers.\cite{SasakiPRL} However, in the
experiment the triplet lifetimes changed by a factor of two only.
The main reason for this difference is probably the ellipticity of
their QDs, which mixes states with different angular momenta and
hence weakens the efficiency of the $\Delta M\!=\!\pm 1$ selection
rule.

In summary, we have estimated the electron TS spin relaxation
rate due to SO coupling in weakly-confined cylindrical
GaAs/AlGaAs QDs.
Experimentally observed trends of TS relaxation in magnetic
fields\cite{FujisawaNATJPCM,SasakiPRL,HansonPRL}
are well understood in terms of the competing SO coupling and
phonon emission efficiency.
Significant differences have been found as compared to the
well-known single-electron spin-flip case, including a critical role
of the dot confinement to determine the phonon emission efficency.
These differences arise from the different orbital initial and final
states, and the (usually larger) transition energies. We predict
very long triplet lifetimes using QD geometries that lead to
suppressed phonon emission. Improved lifetimes can also
be obtained in four-electron QDs by selecting triplet states which
do not fulfill the $\Delta M\!=\!\pm 1$ selection rule.

We are grateful to X. Cartoixa, M. Florescu and F. Troiani for discussions.
We acknowledge support from the Italian Ministry for University and
Scientific Research under FIRB RBIN04EY74, CINECA Calcolo parallelo 2006,
and Marie Curie IEF project MEIF-CT-2006-023797.


\begin{thebibliography}{22}
\bibitem{details_calc}
Well-converged few-electron states are obtained with a basis of 56
single-electron spin-orbitals, which for four electrons may form up
to $\sim 3 \cdot 10^5$ Slater determinants. The SO interaction
Hamiltonian [Eq.~(\ref{eq1})] is diagonalized in a basis of 42
fully-correlated few-electron states.
\bibitem{a0T}
The sharp variation in the relaxation rate of Fig.~\ref{Fig2} as the magnetic
field is switched on is due to removal of the level order imposed by SO interaction
at zero field.\cite{PietilainenPRB} However, in GaAs QDs SO interaction is energetically
small and this level order will already be removed at $B\!=\!0$ by any small deviation from the
perfect circular confinement. Thus, we do not expect it to be observable and do not
discuss it here.
\end{thebibliography}
\end{document}